# HIGHER ORDER MODES HOM'S IN COUPLED CAVITIES OF THE FLASH MODULE ACC39

I.R.R. Shinton[†*], R.M. Jones[†‡], Z. Li[°], P. Zhang[†‡]*
[†]School of Physics and Astronomy, The University of Manchester, Manchester, U.K.
[*]The Cockcroft Institute of Accelerator Science and Technology, Daresbury, U.K.
[°]SLAC, Menlo Park, California, USA
[‡]DESY, Hamburg, Germany

*Abstract*

We analyse the higher order modes (HOM's) in the 3.9GHz bunch shaping cavities installed in the FLASH facility at DESY. A suite of finite element computer codes (including HFSS and ACE3P) and globalised scattering matrix calculations (GSM) are used to investigate the modes in these cavities. This study is primarily focused on the dipole component of the multiband expansion of the wakefield, with the emphasis being on the development of a HOM-based BPM system for ACC39. Coupled inter-cavity modes are simulated together with a limited band of trapped modes.

## INTRODUCTION

At the FLASH facility at DESY a HOM BPM system has been installed and has been successfully used to both align the beam and provide useful cavity diagnostics since 2006 with beam based alignment within 5μm [1].

The accelerating cavities at FLASH (formally TTFII) are based on the TESLA shape. There are two dipole bands below cut-off that are well separated and readily identifiable. This makes them ideal candidates for HOM BPM diagnostic systems. A new module, ACC39 consisting of four third harmonic cavities has recently been installed at FLASH. It flattens the overall electric field profile. The HOMs in the ACC39 cavities are above the cut-off of the connecting beam tubes. As a consequence, these are multi-cavity modes.

Previous experimental studies of both beam and non-beam based measurements [2] have highlighted the difficulty of finding isolated modes suitable for use in a HOM BPM system. This complicates both the experimental analysis and the design of HOM BPM diagnostics. These modes are not readily identifiable since their frequencies will be shifted from the idealised single-cavity 9-cell simulations [3]. As a result ACC39 must be simulated in its entirety. This numerical simulation is computationally expensive in terms of both time and memory resources. All previous numerical results of ACC39 were generated using GSM or CSC [4]. These techniques enabled rapid calculation of large structures through the discretization of the problem into a series of concatenated smaller sections.

Scattering matrices of the entire structure are simulated, matching and misalignment errors are rapidly incorporated within these simulations. However it is difficult to properly obtain the complete field within these coupled structures and in particular to identify trapped modes.

The electromagnetic computational suite ACE3P has been developed by the ACD group at SLAC [5] over the past 20 years, and is capable of large scale parallel computational simulations.

We utilized the frequency domain code OMEGA3P, part of the ACE3P suite, to obtain the modes of the structure. The bellows sections are neglected in these initial simulations in order to obtain rapid results.

## BENCHMARKING OF ACC39

Prior to modelling the complete structure a limited number of cells are simulated. This allows an accurate and rapid comparison with the finite element code HFSS.

The geometry, together with the e-field, displayed in Fig. 1, is a combination of two mismatched ACC39 couplers and two end cells

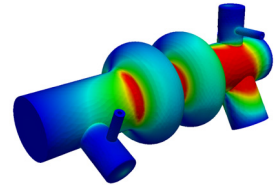

Figure 1: Plot of the magnitude of the electric field taken from OMEGA3P simulation of the resonant frequency at 4.049GHz.

The results from HFSS compared to those from ACE3P in Table 1 indicate reasonable agreement between the independent codes.

Table 1: Comparison of the Kick Factors Predicted by HFSS and OMEGA3P

|      | HFSS  |              | OMEGA3P |              |
|------|-------|--------------|---------|--------------|
| Mode | F:GHz | Kick(V/C/m)  | F:GHz   | Kick(V/C/m)  |
| 1    | 4.049 | $3.24 \times 10^{12}$ | 4.049 | $3.32 \times 10^{13}$ |
| 2    | 4.089 | $2.27 \times 10^{13}$ | 4.088 | $2.31 \times 10^{13}$ |
| 3    | 4.200 | $1.27 \times 10^{13}$ | 4.198 | $1.27 \times 10^{13}$ |

## APPLICATION OF OMEGA3P TO ACC39

The majority of the modes in the cryomodule are able to propagate out through the beam tubes. Here we focus on the properties of the modes which are trapped in either the cavity or the beam-pipe region.

Eigen modes produced with OMEGA3P simulations that are largely restricted to the beam-pipe region are illustrated in Fig. 2, together with an experimental measurement of the transmission through the couplers in which the beam is moved horizontally and vertically. In the data illustrated, radiation from the second cavity, out of the string of eight, was monitored. The field profile for a selected set of modes in this band is illustrated in Fig. 3. It is evident that these are indeed beam-pipe modes. However the degree to which they couple to the beam, represented by R/Q, is rather weak.

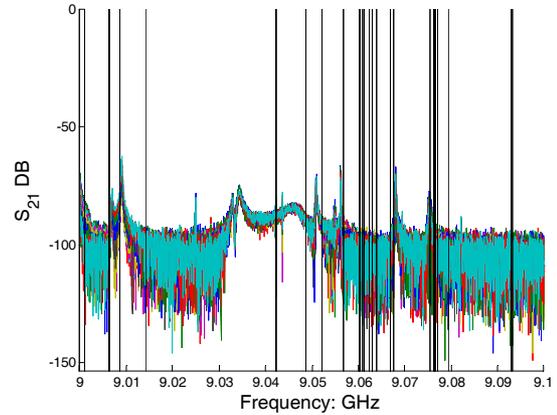

Figure 4: Comparison of 5th dipole band experimental data (in which beam moved with cross offset – various colours relate to different offsets) to OMEGA3P eigenvalues (black lines).

The electric field distribution of some of these modes is displayed in Fig. 5. These modes are not entirely localised to one cavity but are beam-pipe modes and are localised between two cavities. Despite this complicating the experimental analysis they still have the potential to be used as HOM BPM's.

The modes that are truly trapped in the fifth dipole band are in the region of 9.06GHz to 9.065GHz. These modes are predicted by the numerical simulations and displayed in Fig. 6, but possess low R/Q values and do not couple well to the beam. This is confirmed from the experimental measurements in Fig. 4. These trapped modes have limited potential for use as HOM BPM's.

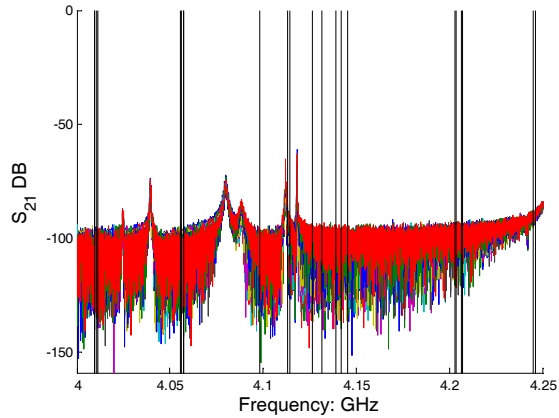

Figure 2: Comparison of 1st beam-pipe mode band experimental data (in which beam moved with cross offset – various colours relate to different offsets) to ACE3P eigen values (black lines).

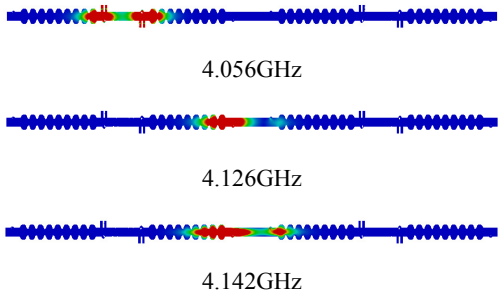

4.056GHz

4.126GHz

4.142GHz

Figure 3: ACE3P magnitude electric field distribution from the 1st beam-pipe mode band for modes.

The next most promising region for a HOM based BPM system is the 5$^{th}$ dipole band, which from the initial idealised 9cell simulations demonstrated that there were potentially trapped modes in this region [4]. Once again we observe that there are several modes predicted by ACE3P that coincide with the experimental data and show a linear dependence in beam offset in Fig. 4.

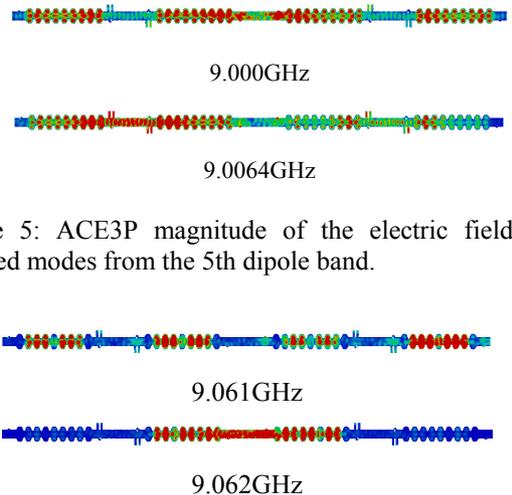

9.000GHz

9.0064GHz

Figure 5: ACE3P magnitude of the electric field for selected modes from the 5th dipole band.

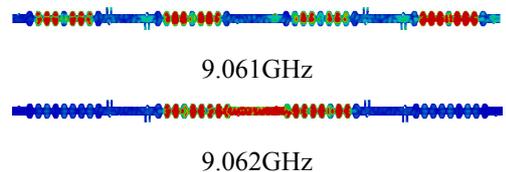

9.061GHz

9.062GHz

Figure 6: Magnitude of electric field from OMEGA3P simulations for the 5$^{th}$ dipole band. The 9.061 GHz mode is a trapped mode and that at 9.062 GHz is an inter-cavity mode.

Suitable HOM BPM diagnostic modes are currently still under investigation. Earlier simulations, based on single cavities, show significant frequency shifts from the multi-cavity results obtained recently with OMEGA3P eigen frequencies. R/Q values obtained with OMEGA3P simulations are shown in Fig. 7.

We note that a region with good coupling exists between 7GHz to 8GHz. This is a region densely populated with multipoles ranging from monopole up to sextupole modes. Mode identification in this region is not straightforward and will be problematic if used for diagnostic purposes.

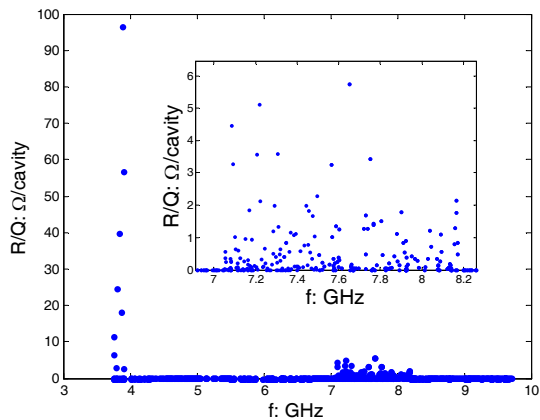

Figure 7: R/Q values calculated using ACDTOOL and OMEGA3P for all modes up 9.8GHz.

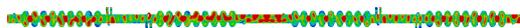

Figure 8: ACE3P magnitude electric field distributions from the multi-cavity mode at 7.301GHz.

The first quadrupole band runs from 6.5GHz to 6.8GHz, some of these modes are indicated in Fig. 9. This band also contains localised modes. However they are unlikely they are unlikely to be relevant as modes for beam diagnostics.

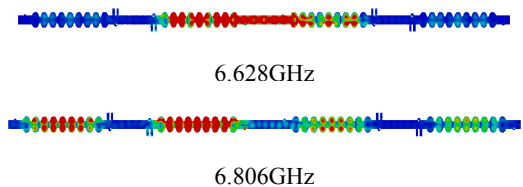

6.628GHz

6.806GHz

Figure 9: Magnitude electric field distributions from simulations with OMEGA3P for the 1st quadrupole band.

## FINAL REMARKS

The beam-pipe modes from either the first beam-pipe band or the fifth dipole band are difficult to use for phase determination. They also have small R/Q values and hence they couple weakly to the beam. The most suitable beam-pipe mode for beam diagnostic purposes is that at 4.126GHz.

The eigen-mode simulations do not predict all experimentally measured modes. OMEGA3P simulations together with the time domain simulations S3P are in progress to identify these modes.

## ACKNOWLEDGEMENTS

This research has received funding from the European Commission under the FP7 Research Infrastructures grant no. 227579. We are pleased to acknowledge E. Vogel [6] for contributing important geometrical details on the cavities and coupler within ACC39.